\documentclass[12pt,a4paper,final,noeprint]{iopart}

\usepackage{iopams}  
\usepackage{graphicx}
\usepackage{float}
\usepackage{physics}
\usepackage{braket}
\usepackage{enumitem}
\usepackage[breaklinks=true,colorlinks=true,linkcolor=blue,urlcolor=blue,citecolor=blue]{hyperref}
\usepackage{caption}
\usepackage{fancyhdr}
\newcommand{\PME}{Pritzker School of Molecular Engineering, University of Chicago, Chicago, IL 60637, USA}

\begin{document}

\title{Nanophotonic quantum network node with neutral atoms and an integrated telecom interface}

\author{Shankar G. Menon}
\address{\PME}
\author{Kevin Singh}
\address{\PME}
\author{Johannes Borregaard\footnote{jborregaard@math.ku.dk}}
\address{QMATH, Department of Mathematical Sciences, University of Copenhagen, Copenhagen, Denmark}
\author{Hannes~Bernien\footnote{bernien@uchicago.edu}}
\address{\PME}

\begin{abstract}

The realization of a long-distance, distributed quantum network based on quantum memory nodes that are linked by photonic channels remains an outstanding challenge. We propose a quantum network node based on neutral alkali atoms coupled to nanophotonic crystal cavities that combines a long-lived memory qubit with a photonic interface at the telecom range, thereby enabling the long-distance distribution of entanglement over low loss optical fibers. We present a novel protocol for the generation of an atom-photon entangled state which uses telecom transitions between excited states of the alkali atoms. We analyze the realistic implementation of this protocol using rubidium and cesium atoms taking into account the full atomic level structure and properties of the nanophotonic crystal cavity. We find that a high fidelity entangled state can be generated with current technologies.

\end{abstract}

\section{Introduction}

Quantum networks have been envisioned as the underlying platform for revolutionizing technologies including secure communication~\cite{Gisin2002,Wehner2018}, distributed quantum computing~\cite{Nickerson2013,Monroe2014}, and quantum enhanced metrology~\cite{Proctor2018,Gottesman2012,Emil2019}. At the same time, quantum networks provide avenues for studying fundamentals of nature such as entanglement over large distances~\cite{Hensen2015}.
Elementary quantum networks with optical channels and matter qubits have been demonstrated in multiple systems such as ions~\cite{Moehring2007,Hucul2015}, atoms~\cite{Ritter2012,Hofmann2012,Yu2019}, and atom-like systems in solids~\cite{Bernien2013,Bhaskar2019,Delteil2016,Stockhill2017,Usmani2012}. However, all demonstrations so far have been restricted to two nodes and limited distances due to low entanglement rates or short coherence times of the qubits. 

In order to increase entanglement rates and the distance between network nodes, efficient light-matter interfaces and low-loss optical channels are required. While free space optical links can provide low loss channels~\cite{Ursin2007,Yin2017} they are hard to implement in metropolitan settings and face challenging atmospheric conditions~\cite{Miao2005,Liao2017}. Low-loss optical fibers exist in the telecom communication band and decades of research and development have led to losses as low as 0.1419 dB/Km~\cite{Tamura2018}.
Quantum networks can leverage this technology by developing network nodes with high bandwidth, long coherence times, and telecom wavelength operation.

\begin{figure}
    \centering
    \includegraphics[width = 14cm]{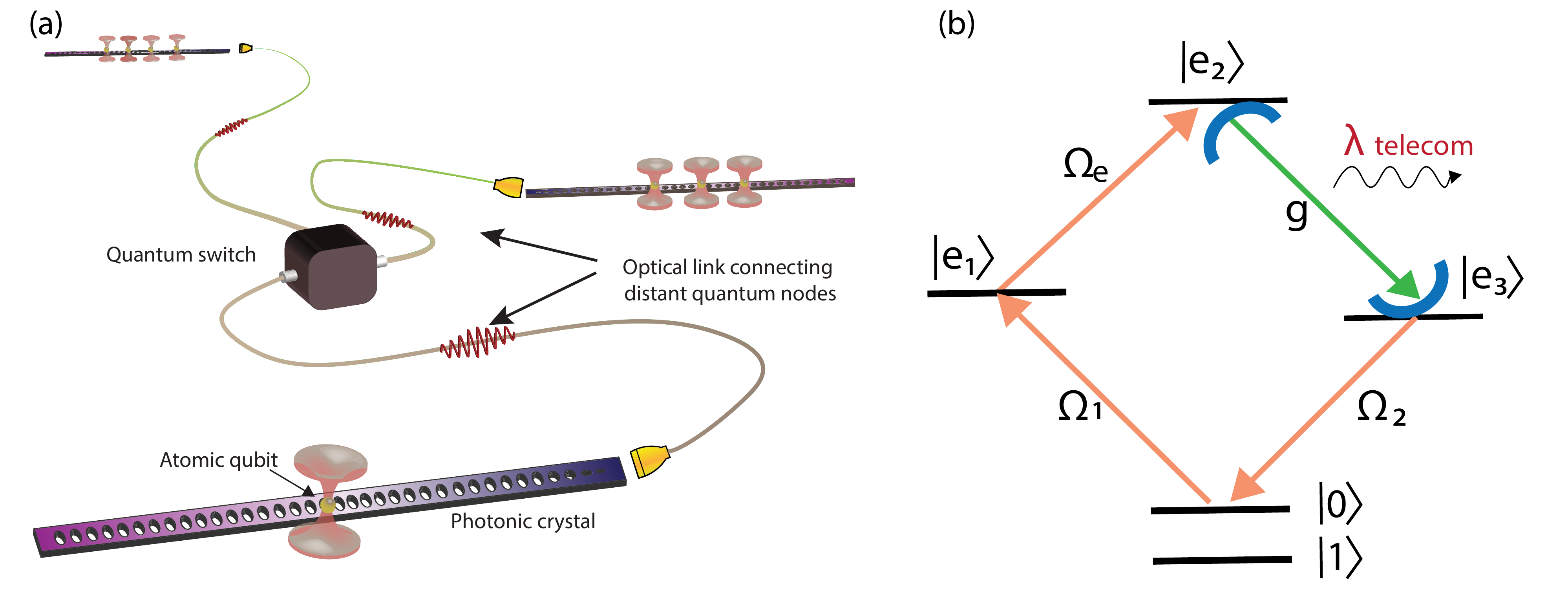}
    \caption{(a) Illustration of a quantum network based on neutral atoms. Here each node consists of an atom coupled to a photonic crystal cavity, with clock states as the qubit states. Distant nodes are linked via telecom optical fibers. (b) Relevant atomic levels of the diamond scheme. $ \ket{0}$ and $\ket{1}$ are the qubit states. Starting from $\ket{0}$, a  pulse $\Omega_1$ takes the electron to $\ket{\textrm{e}_1}$. From there a constant field $\Omega_{\mathrm{e}}$ excites the electron to $\ket{\textrm{e}_2}$. Subsequently a single telecom photon is emitted into the cavity and coupled to the optical fiber by decaying to $\ket{\textrm{e}_3}$. Finally, a calibrated pulse $\Omega_2$ takes the electron back to $\ket{0}$.}
    \label{fig:figure1}
\end{figure}

Atomic qubits, including atom-like systems in solids, are promising candidates for quantum network nodes (see figure~\ref{fig:figure1}(a)) as they combine long coherence time and quantum control with optical transitions. Neutral atoms are particularly attractive as they are indistinguishable and provide highly coherent qubit states and a coherent photonic interface. Individual neutral atoms can be trapped using optical tweezers~\cite{Schlosser2001} and methods of scaling this approach to arrays of atoms have recently become available~\cite{endres2016,Barredo2016,Kim2016,Brown2019}. Efficient light-matter interfaces to these qubits can be established using optical cavities~\cite{Reiserer2015}. Many important steps in achieving a quantum network with neutral atoms have been demonstrated including quantum memory~\cite{Specht2011}, atom-photon quantum gates~\cite{Reiserer2014,Tiecke2014}, and two-node entanglement~\cite{Ritter2012,Hofmann2012}. However, telecom operation of atomic network nodes remains challenging as nearly all atomic species that are compatible with laser trapping and cooling do not have telecom wavelengths transitions from the ground state. While frequency conversion can be employed to shift emitted photons from the near-infrared and visible to the telecom range~\cite{Yu2019,Tchebotareva2019}, this process adds noise and has finite conversion efficiency. Therefore, it can limit both the rate of entanglement generation and the fidelity of the entangled states.

Here, we propose a fiber-based quantum network with individual nodes of neutral alkali atoms coupled to a nanophotonic crystal cavity (PCC) (see figure~\ref{fig:figure1}(a)) and a multilevel excitation protocol that yields emission in the telecom range.
We show that this node is capable of generating a high fidelity atom-photon entangled state, which is the essential functionality required for distributing entanglement. The protocol is robust under realistic conditions and we evaluate the performance including accurate simulations of the full atomic level structure, the nanophotonic cavity design and its effect on the polarization purity of laser excitation pulses. Compared to previous neutral atom proposals based on crossed, macroscopic cavities~\cite{Uphoff2016} or ytterbium coupled to a nanophotonic crystal cavity~\cite{Covey2019}, this protocol presents an alternative that is compatible with well-controlled alkali `workhorse' atoms such as rubidium and cesium and only requires a single nanophotonic cavity. 
We establish the experimental feasibility of our protocol and show that it is within reach of current technology. 

In this article, we will first discuss the general scheme and implementation of the atom-telecom photon entanglement generation with analytical and numerical estimates of the error scaling. This is followed by the concrete implementation using alkali atoms, in particular rubidium and cesium, together with numerical estimates of the error scaling. We then discuss the design and performance of a suitable telecom photonic crystal cavity and finally demonstrate the feasibility of creating a high fidelity atom-photon entangled state by analyzing the performance of the protocol for the full system.

\section{The protocol}
 We consider the generic diamond level scheme~\cite{Chaneliere2006,Willis2009} shown in figure~\ref{fig:figure1}(b), which captures the general structure of alkali atoms such as rubidium and cesium. The scheme has two parts; a qubit part consisting of ground spin states $\ket{0}$ and $\ket{1}$ and a photon-generation part consisting of excited spin states $\ket{\textrm{e}_1}$, $\ket{\textrm{e}_2}$, and $\ket{\textrm{e}_3}$. To generate spin-photon entanglement, the qubit states are initially prepared in a superposition state ($\ket{0}$ + $\ket{1}$)/$\sqrt{2}$. A laser-driven two-photon transition then transfers the population in the $\ket{0}$ state to the excited state $\ket{\textrm{e}_2}$, from which the atomic state coherently relaxes to state $\ket{\textrm{e}_3}$ by emitting a telecom cavity photon. Finally, a strong laser pulse transfers the population from $\ket{\textrm{e}_3}$ back to the ground state $\ket{0}$. At the end of this cycle, the qubit states are flipped ($\ket{0}\leftrightarrow\ket{1}$) by a $\pi$ pulse on the qubit transition and the cycle is repeated. Ideally, this creates the state

\begin{equation} \label{eq:idealstate}
    \ket{\psi} = \frac{1}{\sqrt{2}}(\ket{0}\ket{\lambda_{\mathrm{L}}} + \ket{1}\ket{\lambda_{\mathrm{E}}} ).
\end{equation}
Here $\ket{\lambda_{\mathrm{E}}}$ ($\ket{\lambda_{\mathrm{L}}}$) represents an early (late) telecom photon. This is a maximally entangled state between the atomic qubit and a photonic qubit in a time-bin encoding, which can be used to distribute entanglement between distant atoms through photonic Bell measurements~\cite{Simon2003}. 

The ideal evolution described above is in the absence of atomic spontaneous emission and cavity loss. To estimate how such imperfections limit the quality of the spin-photon entanglement, we analytically estimate the dynamics of the scheme. In a suitable rotating frame, the coherent dynamics are governed by the Hamiltonian
\begin{equation} \label{eq:Hamil1}
\hat{H}=\Omega_1(t)\ket{\textrm{e}_1}\bra{0}+\Omega_{\mathrm{e}}\ket{\textrm{e}_2}\bra{\textrm{e}_1}+\Omega_2(t)\ket{0}\bra{\textrm{e}_3}+g\ket{\textrm{e}_3}\bra{\textrm{e}_2}\hat{c}^\dagger +h.c.,    
\end{equation}
 where $h.c.$ denotes the Hermitian conjugate of the displayed expression and $\hat{c}$ is the annihilation operator of the cavity field. The atom-cavity coupling is characterized by $g$ while atom-laser couplings are characterized by $\Omega$'s. Note that $\Omega_1,$ $ \Omega_2$ are time dependent while $\Omega_{\mathrm{e}}$ is constant. We have assumed all laser frequencies ($\omega_{\mathrm{L1}}, \omega_{\mathrm{Le}}$, and $\omega_{\mathrm{L2}}$) to be resonant with the relevant atomic transitions and that $\omega_{\mathrm{cav}}+\omega_{\mathrm{L2}}=\omega_{\mathrm{L1}}+\omega_{\mathrm{Le}}$ where $\omega_{\mathrm{cav}}$ is the cavity resonance frequency. We choose to have resonant driving frequencies because we want to transfer population quickly between the excited levels to circumvent spontaneous emission. 
 
Spontaneous emission is described by Lindblad operators
 $\hat{L}_1=\sqrt{\gamma_1}\ket{d_1}\bra{\textrm{e}_1}$, $\hat{L}_2=\sqrt{\gamma_2}\ket{d_2}\bra{\textrm{e}_2}$, and $\hat{L}_3=\sqrt{\gamma_3}\ket{d_3}\bra{\textrm{e}_3}$. Here, the decay rates are denoted by $\gamma$'s and we have introduced \emph{dump}-levels ($\ket{d_i}$), which allow us to disregard the evolution of the system following spontaneous emission from one of the excited spin states. This amounts to a worst-case scenario where we assume that any spontaneous emission brings the system to a state with zero overlap with the desired target state in Eq.~(\ref{eq:idealstate}). 

We can analytically solve the equations of motion assuming the atom is initially in state $\ket{0}$ (the dynamics are trivial if the atom is in state $\ket{1}$) with zero cavity photons. We describe the decay of the cavity field through a Lindblad operator $\hat{L}_{\mathrm{c}}=\sqrt{\kappa}\hat{c}$, where $\kappa=\kappa_{\mathrm{f}}+\kappa_{\mathrm{l}}$. Here, $\kappa_{\mathrm{f}}$ describes the desired outcoupling to a fiber while $\kappa_{\mathrm{l}}$ describes intra-cavity loss. Note that such loss is detectable and thus only affects the efficiency of the protocol. Adopting a stochastic wave function approach~\cite{Jean1992}, we solve for the no-jump evolution of the system described by the non-Hermitian Hamiltonian $H_{\mathrm{NJ}}=\hat{H}-(\mathrm{i}/2)\sum_{i} \hat{L}^{\dagger}_i\hat{L}_i$. We consider $\Omega_1(t)$ and $\Omega_2(t)$ to be square pulses such that $\Omega_1(t)=\Omega'_1$ is constant for $t\leq t_1$ and zero otherwise while $\Omega_2(t)=\Omega'_2$ is constant for $t_1\leq t\leq t_2$ and zero otherwise. 
This allows us to find the population $\rho_{\mathrm{0,\lambda}}$ in state $\ket{0}\ket{\lambda_{\mathrm{E}}}$ at the end of the sequence as detailed in~\ref{ap:theory}.

\begin{figure}
    \centering
    \includegraphics[width = 14cm]{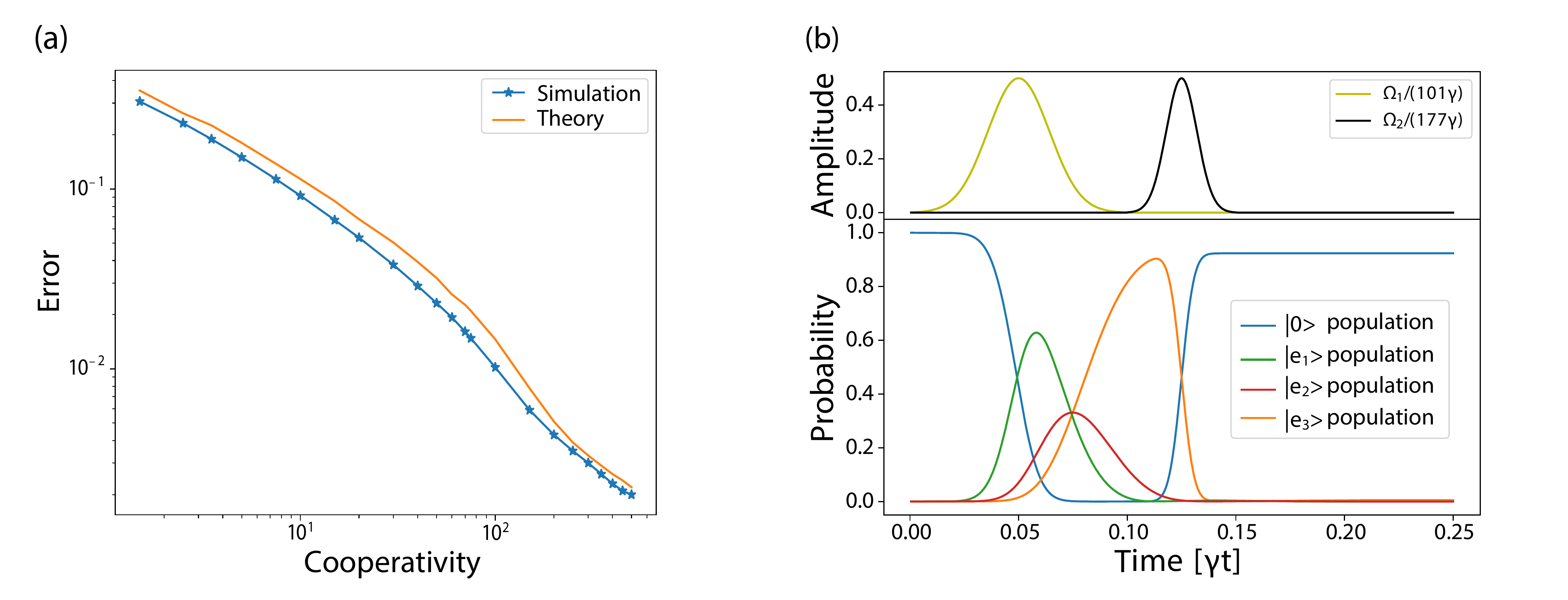}
    \caption{(a) Error scaling with cooperativity for the analytical estimate (orange) and the numerical simulation (blue markers); the solid line for simulation is a guide to the eye. The simulation is in agreement with the analytical prediction of the error scaling. (b) Time evolution of the excitation pulses (top) and probability in different states (bottom) for a cooperativity of $c=10$ and optimized laser couplings. The first pulse $\Omega_1$ (yellow) depopulates $\ket{0}$ (dark blue). The presence of light field $\Omega_{\mathrm{e}}$ and the cavity leads to a transfer to state $\ket{\textrm{e}_3}$ (orange). After the photon has been emitted from the cavity, on timescale $1/\kappa$, the resulting state is $\ket{\textrm{e}_3,0}$. A pulse $\Omega_2$ takes the population built in $\ket{\textrm{e}_3}$ to $\ket{0}$. In both plots $\gamma_1=\gamma_2=\gamma_3=\gamma$ and $\kappa=2000\gamma$. 
    }
    \label{fig:error-scaling}
\end{figure}

Due to the symmetric nature of the excitation scheme before and after the qubit flip, the target state in Eq.~(\ref{eq:idealstate}) can be estimated as $F\approx\kappa\rho_{0,\lambda}/\kappa_{\mathrm{f}}$. Choosing $\Omega'_1\sim C\gamma_1$, $\Omega_{\mathrm{e}}\sim C\gamma_2$, and $\gamma_1 t_1\sim\ln(C)/C$, we find an error scaling of
\begin{equation} 
1-F\approx \frac{\ln(C)}{C},
\end{equation}
assuming 
$\Omega'_2\gtrsim C\gamma_3$. Here, we have defined the cooperativity $C=\frac{g^2}{\kappa(\gamma_2+\gamma_3)}$. Notably, the error quickly decreases with the cooperativity allowing fidelities $>90\%$ for modest cooperativities of $C\gtrsim10$ (see figure~\ref{fig:error-scaling}(b)). Furthermore, we note that some errors which are currently included in our fidelity estimate such as spontaneous decay from the $\ket{\textrm{e}_1}$ and $\ket{\textrm{e}_2}$ state may, in principle, be detected by the absence of a telecom-photon. Employing such error-detection can therefore further boost the fidelity at the expense of a slight decrease in success probability of the protocol.   


The analytical estimate of the error scaling is obtained by adiabatically eliminating certain coherences as discussed in~\ref{ap:theory}. To validate the estimate and further minimize the error, we numerically simulate the dynamics by solving the full master equation. The result of such a simulation is shown in figure~\ref{fig:error-scaling}(b). The simulation allows us to track the transfer of the initial population in state $\ket{0}$ during the photon-generating loop.  
The fidelity is defined as the population that reaches the initial state ($\ket{0}$) after a complete pulse sequence. 
We optimize the fidelity for a given cooperativity by numerically finding optimal laser pulses
(see figure~\ref{fig:error-scaling}(b)). The minimal error found in the numerical simulation agrees with the error scaling predicted by the analytical estimate (see figure~\ref{fig:error-scaling}(a)). 

\section{Implementation with alkali atoms}

The generic level structure considered so far has to be implemented in real atoms. We consider cesium and rubidium atoms since they have telecom transitions the meta-stable first excited states and can be trapped near a PCC~\cite{Thompson2013b,Goban2014,Samutpraphoot2019,Kim2019} with strong light-matter coupling~\cite{Samutpraphoot2019}. Here, we focus on an implementation with cesium atoms while a similar implementation with rubidium atoms is discussed in \ref{ap:level scheme}. 

\begin{figure}
    \centering
    \includegraphics[width = 14cm]{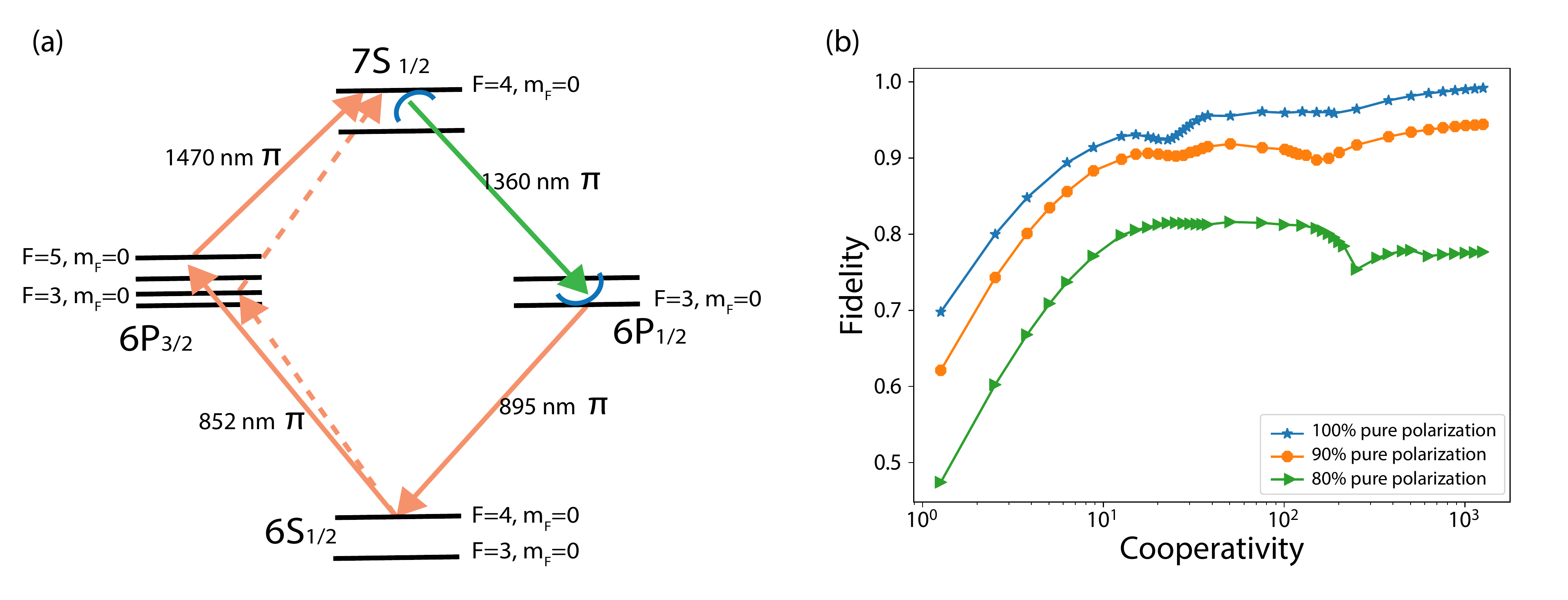}
    \caption{(a) A realistic implementation of the diamond scheme based on the cesium level structure. Here Zeeman sub-level states $\mathrm{m_F}=0$ of $\ket{6S_{1/2}, \mathrm{F}=4}$, $\ket{6P_{3/2},  \mathrm{F}=5}$, $\ket{7S_{1/2}, \mathrm{F}=4}$, $\ket{6P_{1/2}, \mathrm{F}=3}$ act as $\ket{0}$, $\ket{\textrm{e}_1}$, $\ket{\textrm{e}_2}$, and $\ket{\textrm{e}_3}$, respectively. A transition to $\ket{6P_{3/2},  \mathrm{F}=3, \mathrm{m_F}=0}$ is allowed by selection rules, but it is off-resonant. All transitions are $\pi$ polarized. (b) Scaling of the maximum fidelity with cooperativity for polarization purities of 100\%, 90\%, and 80\% for all the transitions. Solid lines are a guide to the eye. Equal contribution from $\sigma^+$ and $\sigma^-$ are considered in case of polarization impurity. All hyperfine sub-levels and Zeeman sub-levels are considered in the simulation. The dips in the graph correspond to off-resonant couplings leading to lower fidelity. The simulation variables are parameterized based on the decay rate ($\gamma$) from the $7S_{1/2}$ state given by $2 \pi\times3.28$ MHz. The decay rate from $6P_{3/2}$ is $2 \pi\times 5.2$ MHz and $6P_{1/2}$ is $2 \pi\times4.6$ MHz corresponding to $1.58\gamma$ and $1.38\gamma$ respectively. The cavity decay rate $\kappa$ was assumed to be $200\gamma$. }
    \label{fig3}
\end{figure}

Cesium has telecom transitions to $7S_{1/2}$ from both $6P_{1/2}$ and $6P_{3/2}$ with wavelengths 1360~nm (O band) and 1470~nm (S band) respectively.
These levels are split into hyperfine levels and further into Zeeman sub-levels in the presence of magnetic fields. As a result, the atomic level structure includes many more levels than required for the protocol. Nonetheless, by means of selection rules, laser frequencies, and a suitable choice of qubit states, we can realize the diamond level scheme as shown in figure~\ref{fig3}(a). We choose magnetic field insensitive clock states, $\ket{6S_{1/2}, \mathrm{F}=4, \mathrm{m_F}=0}$, $\ket{6S_{1/2}, \mathrm{F}=3, \mathrm{m_F}=0}$, as our qubit states $\ket{0}$ and $\ket{1}$, respectively. The superposition of the clock states can be created by first initializing in the state $\ket{6S_{1/2}, \mathrm{F}=4, \mathrm{m_F}=0}$ via optical pumping and then applying a $\pi/2$ pulse between the clock states. 

We choose $\pi$-polarized laser excitation pulses, which allow us to make use of the selection rules $\Delta \mathrm{F}=0,\pm 1$ and $\Delta \mathrm{m_F}=0$ to realize the photon-generation part of the level scheme. We identify $\ket{6P_{3/2}, \mathrm{F}=5, \mathrm{m_F}=0}$ as $\ket{\textrm{e}_1}$, $\ket{7S_{1/2}, \mathrm{F}=4, \mathrm{m_F}=0}$ as $\ket{\textrm{e}_2}$ and $\ket{6P_{1/2}, \mathrm{F}=3, \mathrm{m_F}=0}$ as $\ket{\textrm{e}_3}$ respectively. The first pulse takes the population to $\ket{6P_{3/2}, \mathrm{F}=5, \mathrm{m_F}=0}$ state. Depending on the strength and spectrum of the pulse, it will also off-resonantly drive the transition to $\ket{6P_{3/2}, \mathrm{F}=3, \mathrm{m_F}=0}$ state, but part of this is coupled back into the cycle by $\Omega_{\mathrm{e}}$. The transitions with $\Delta \mathrm{F}=0$ have vanishing Clebsch-Gordan coefficients when $\Delta \mathrm{m_F}=0$. This selection rule blocks all other off-resonant transitions in the scheme. 

The above implementation in cesium is conditioned on polarization selection rules and is thus sensitive to polarization fluctuations. Even a small percentage of erroneous polarization will result in coupling to other levels 
and a high polarization purity is imperative for the implementation of the scheme. As we will see in the following section, the presence of nanostructures can have considerable effects on the polarization of a beam in its vicinity. To understand the effects of impure polarization, we simulate the full atomic level structure and consider small admixtures of $\sigma^+$ and $\sigma^-$-polarized light along with the intended $\pi$ polarized driving pulses. The effect of an imperfect polarization purity on the fidelity is shown in figure~\ref{fig3}(b). 
As expected, the fidelity is limited by  the presence of other polarizations for high cooperativities. 
The coupling to other hyperfine levels also causes a non-trivial dependence on the cooperativity with local minima appearing when $\Omega$ or the cavity coupling g become comparable to the hyperfine splitting. 

\section{Nanophotonic crystal cavity and polarization purity}

\begin{figure}
    \centering
    \includegraphics[width = 14cm]{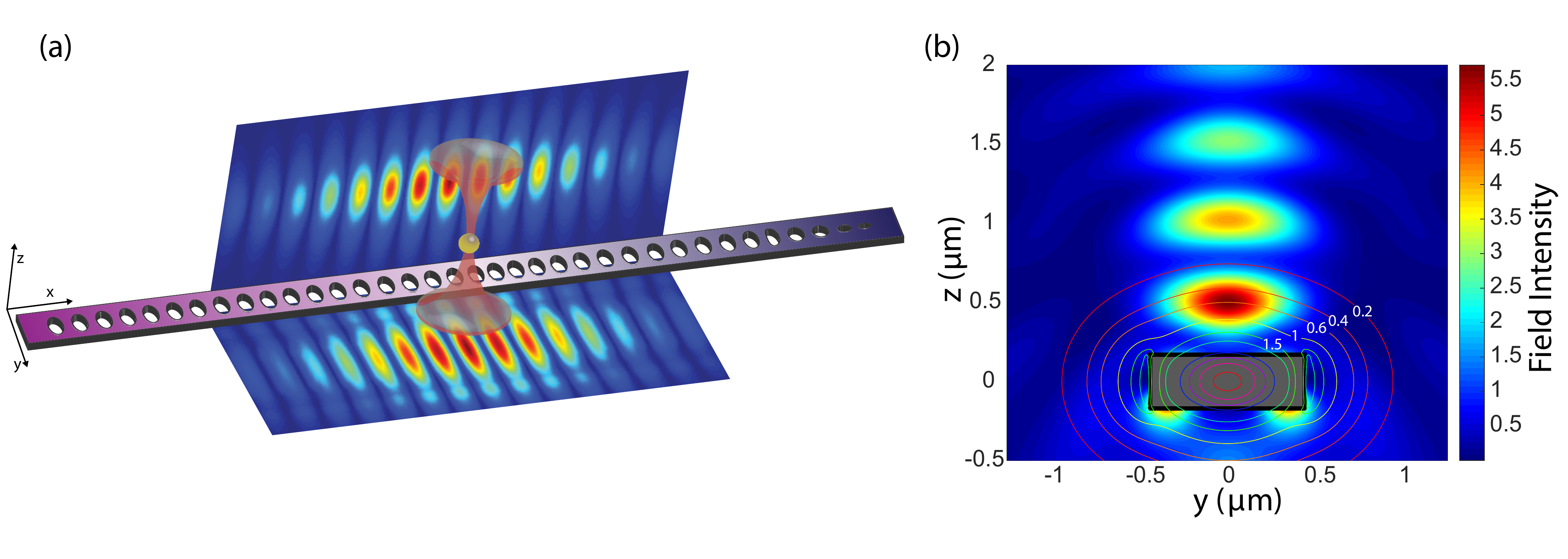}
    \caption{(a) Design of a photonic crystal cavity along with the resonant mode profile at 1360~nm corresponding to the transition between $7S_{1/2}$ and $6P_{1/2}$ of cesium. (b) Standing wave formation on a waveguide of width 876~nm and thickness of 340 nm when an optical tweezer beam at 910~nm is incident from above. Here the incident gaussian is normalized at a distance of $\sim 2$~$ \mu$m from the cavity. A cross section of the waveguide is shown (dark rectangle). The atom can be trapped in the electric field maxima closest to the surface. The contours show lines of constant atom-cavity coupling g (in units of $2\pi\times$~GHz) exponentially decaying from the center of the waveguide. Assuming critical coupling, i.e $\kappa_{\mathrm{f}}=\kappa_{\mathrm{l}}$, the corresponding cooperativities are in the range from 5 to 260.}
    \label{fig4}
\end{figure}

\begin{figure}
    \centering
    \includegraphics[width = 14cm]{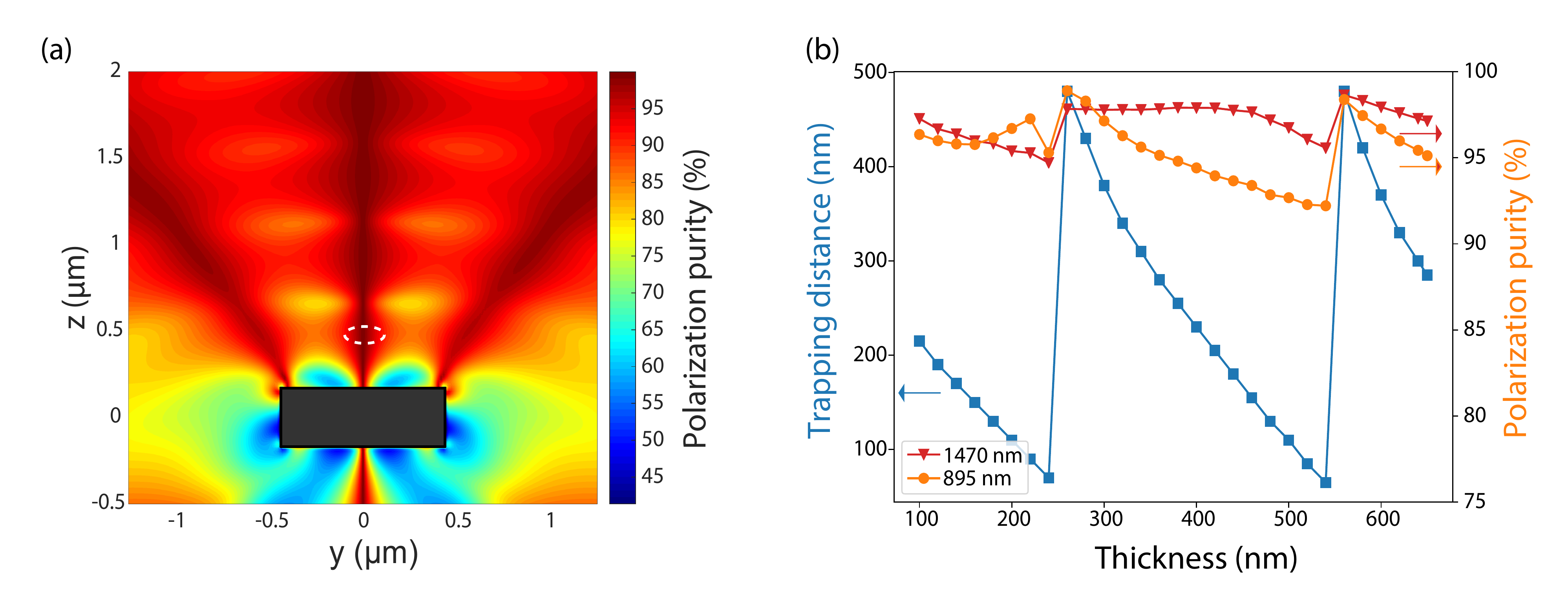}
    \caption{(a) A nanostructure modifies the polarizations of an incident light field near its surface. An 895 nm plane wave with polarization in the y direction is incident on the waveguide (black rectangle) from the positive z axis. The time averaged polarization purity around the waveguide is shown in the cross section. The region where the atom would be trapped based on the intensity distribution in figure~\ref{fig4}(b) is marked by a dashed white ellipse. (b) The trapping distance (blue squares) of the atom from the surface of the device can be modified by changing the thickness of the waveguide. The corresponding polarization purity also changes with thickness of the waveguide (orange dots for 895~nm and red triangles for 1469~nm). The polarization purity is averaged over an area of 200$\times$200$\times$50 $\text{nm}^3$. Solid lines are a guide to the eye. }
    \label{fig5}
\end{figure}

We design a PCC operating at telecom wavelengths to evaluate the expected cooperativities and effects on the polarization, and thereby the fidelity of the above scheme, using Finite-Difference Time-Domain (FDTD) simulations. PCCs have considerable flexibility in tailoring the atom-photon interaction through high quality factors, small mode volumes, and the ability to deterministically couple multiple atoms~\cite{Tiecke2014,Thompson2013b,Goban2014,Samutpraphoot2019}. 
We design a cavity resonant at 1360 nm corresponding to the transition between $7S_{1/2}$ and $6P_{1/2}$ at 340 nm thickness with a quality factor of $2\times 10^5$ and a mode volume of 0.7 $(\lambda/n)^3$ as shown in figure~\ref{fig4}(a). Here $n$ is the refractive index of the PCC material which is taken here to be 2.016, corresponding to that of commercially available silicon nitride. The contours of the atom-cavity coupling g are shown in figure~\ref{fig4}(b). \\
The atom is coupled to the resonant mode of the cavity through the evanescent field which decays exponentially with increasing distance from the surface of the structure. We consider trapping an atom in the standing wave formed by an optical tweezer on the surface of a waveguide~\cite{Thompson2013b}. Figure~\ref{fig4}(b) shows this standing wave formation when the incident tweezer beam interferes with the reflected beam from the waveguide. An atom can be loaded in the trap formed by the intensity maxima closest to the cavity by first loading the atom into the tweezer from an magneto-optical trap and then adiabatically moving the tweezer to the cavity surface~\cite{Thompson2013b}. Assuming critical coupling with a quality factor of 2 $\times$ $10^5$, cooperativities of 50 can be achieved at the region of trapping which will allow high fidelity implementation. The assumption of critical coupling will, however, reduce the collected photon number from the cavity to half, which would affect the success probability but not the fidelity of the protocol.\\
We now look at how the polarization purities of the addressing beams are affected by the presence of this cavity. The classical light fields $\Omega_1$, $\Omega_{\mathrm{e}}$, and $\Omega_2$ can address the atom based on the direction set by the quantization axis and the polarization requirement. Since we are interested in $\pi$ polarizations for the implementation of our scheme, we fix the cavity polarization axis (y axis based on the simulated fundamental TE mode) to be our quantization axis. The requirement of $\pi$ (set to be y) polarization for classical fields ($\Omega$) restricts their propagation direction to the x or z axis. With the propagation direction fixed along the z axis, we simulate how the presence of the PCC affects the polarization purity of the exciting beams near the PCC surface. Figure~\ref{fig5}(a) shows the polarization purity of an 895~nm (corresponding to the $\Omega_1$ field) y polarized plane wave incident normal to the surface of the PCC. We find that the reflection from the PCC can have considerable effects on the polarization purity of the driving pulses depending on the position where the atom is trapped. While being confined by the standing wave optical trap, the atom still has some thermal motion and therefore samples over a region determined by the trap geometry and the temperature of the atom~\cite{Samutpraphoot2019}. This leads to an effective average polarization purity as well as an average cooperativity for a given atomic temperature. Positioning the atom in a region of high polarization purity is thus imperative for high fidelity implementation of the scheme.

The position of the atom close to the surface of the cavity can be tuned by changing the thickness of the cavity and thereby changing the trap position~\cite{Thompson2013b}. This tunability allows us to have control over the effective polarization purity as well as the atom's coupling to the cavity field. The trapping distance from the surface and the corresponding polarization purity in the trapping region for different thicknesses of the cavity is shown in figure~\ref{fig5}(b) for a trapping field at 910 nm. In general, we find that polarization impurities are higher when the atom is trapped closer to the cavity.
The given polarization purity is an average over a volume of $200\times200\times50$ $\text{nm}^3$ in the x, y, and z dimensions similar to observed sizes of atomic wavefunctions near a PCC~\cite{Samutpraphoot2019}.

Along with the polarization purity, the trapping distance  also determines the atom's coupling to the cavity field. The effective coupling experienced by the atom is directly proportional to the electric field of the cavity mode at the position of the atom~\cite{Kimble1998}. This electric field falls off exponentially with the distance from the surface. Consequently, strong coupling requires the atom to be placed as close to the cavity as possible. This leads to an apparent trade off between polarization purity which increases with trapping distance and the atom-cavity coupling which decreases with trapping distance.

\section{Combined results}
To estimate realistic fidelities attainable for the diamond scheme, we simulate a cavity with a thickness of 340~nm, quality factor of 2 $\times$ $10^5$, and extract cooperativity and polarization purity at the trapping region from the FDTD simulations. Assuming critical coupling such that $\kappa_{\mathrm{f}} =\kappa_{\mathrm{l}}$, we attain a cooperativity of 35 averaged over the region where the atom would be trapped. At this trapping region we find polarization purities of 93.89\% for 852~nm, 94.99\% for 895~nm, and 97.79\% for 1470~nm plane waves incident from the positive z axis. A polarization purity of 97.66\% was found for the cavity field. Based on these polarization purities, a fidelity of 0.93 can be achieved for the simulated average cooperativity of 35 with square pulses of length 2.5~ns for $\Omega_1$ and 0.5~ns for $\Omega_2$. Cavities of similar and higher quality factors have been made and pulses as short as 50~ps can be made using commercially available electro optic modulators enabling realistic high fidelity implementation of the scheme. Moreover, fidelities greater than 0.9 can be achieved with cooperativities as small as 15 for the simulated polarization purity implying the possibility of high fidelity implementation even in lossy cavities and higher temperatures of the atom. Additionally, further optimization of the cavity can be performed to obtain higher atom-cavity coupling, higher polarization purity, and lower cavity losses to improve the fidelities.

\section{Summary and Outlook}

A future distributed global quantum network will rely on quantum nodes with long coherence times, high bandwidths, and telecom interfaces. We have proposed and analysed the performance of a node architecture that consists of an individual atom coupled to a nanophotonic crystal cavity which operates at telecom wavelengths. We have shown that this node is capable of generating a high fidelity atom-photon entangled state under realistic conditions taking both the full atomic level structure and the properties of the nanophotonic cavity into account. We have analysed the scaling of the fidelity with the cooperativity of the atom-cavity system and find that fidelities larger than 0.9 are attainable with current state of the art technologies. 

Our assumptions are rather conservative and we expect that even higher fidelities are achievable. For instance, lower temperatures of the atom would reduce the region over which the polarizations of the driving fields are averaged and would lead to higher polarization purities. Furthermore, a careful cavity design could enlarge the regions of high polarization purity and align them with the trapping region. Additionally, based on recent results for strong coupling of atoms to nanophotonic crystal cavities at 780~nm, 
we anticipate that cavities with higher quality factors can be fabricated at longer wavelengths~\cite{Asano2017}. 

Finally, the proposed node architecture can be extended in several important directions. A multi-qubit node can be achieved by coupling multiple atoms to the same cavity~\cite{Sorensen2003,Duan2005,Borregaard2015}. Operations between these qubits can be enabled via photon mediated gates through the cavity. Such an architecture will enable advanced network functionalities such as entanglement purification~\cite{Bennett1996,Kalb2017} and quantum repeater protocols~\cite{Uphoff2016,Briegel1998,Childress2006,Borregaard2015pra}, thereby paving the way towards large scale multi-node quantum networks.

\section*{Acknowledgments}

We thank Mikhail Lukin, Alan Dibos, and Jordan Kemp for insightful discussions. JB acknowledges support from Villum Fonden via the QMATH Centre of Excellence (grant no. 10059).

\newpage

\bibliographystyle{naturemag}

\bibliography{refs,telecom}

\section*{Appendix}
\appendix

\section{Analytical estimates} \label{ap:theory}
Here we describe how to analytically solve for the dynamics of the scheme and estimate the performance of the spin-photon entanglement generation. 
Our starting point is the Hamiltonian in Eq.~(\ref{eq:Hamil1}). We assume the atom is initially in state $\ket{0}$ and that the cavity field is vacuum ($\ket{0_{\mathrm{ph}}}$). We focus on the no-jump evolution of the state vector 
\begin{equation}
\ket{\phi}=c_0\ket{0}\ket{0_{\mathrm{ph}}}+c_+\ket{+}+c_-\ket{-}+c_2\ket{\textrm{e}_3}\ket{1_{\mathrm{ph}}},
\end{equation}
subject to the non-Hermitian Hamiltonian $H_{\text{NJ}}=\hat{H}-(\textrm{i}/2)\sum_i \hat{L}^{\dagger}_i\hat{L}_i$ with Lindblad operators as defined below Eq.~(\ref{eq:Hamil1}). We have defined the states $\ket{\pm}=(\ket{\textrm{e}_1}\ket{0_{\text{ph}}}\pm \ket{\textrm{e}_2}\ket{0_{\text{ph}}})/\sqrt{2}$. Note that the overall population decay corresponds to the atom spontaneously decaying to one of the dump levels or that the cavity photon in state $\ket{\textrm{e}_3}\ket{1_{\text{ph}}}$ couples out into fiber ($\kappa_{\mathrm{f}}$) or is lost ($\kappa_{\mathrm{l}}$). Explicitly including the population buildup in these states would correspond to a consistent and probability preserving description of the dynamics. However, since we are interested in a lower bound for the performance of the scheme, we simply view any decay to a dump level as an undetectable decay into a state with zero overlap with the desired target state (Eq.~(\ref{eq:idealstate})). This allows us to neglect the dynamics following such a decay. Furthermore, we first solve for the dynamics for a time period $0\geq t\geq t_1$ where we assume that $\Omega_1(t)=\Omega'_1$ is constant and $\Omega_2=0$. We can therefore also neglect the dynamics following the outcoupling or loss of the cavity photon and simply view this as a build up of population in the state $\ket{\textrm{e}_3}\ket{0_{\text{ph}}}$ (which decays with rate $\gamma_3$).    

From the Schr\"odinger equation ($\frac{\partial}{\partial t}\ket{\phi}=-\mathrm{i}\hat{H}_{\text{NJ}}\ket{\phi}$), we obtain equations of motion for the amplitudes $c_0$, $c_{\pm}$, and $c_2$. Assuming that $\Omega_1\ll\sqrt{2}\Omega_{\mathrm{e}}$, we can adiabatically eliminate the $c_{\pm}$ amplitudes and obtain
\begin{eqnarray}
\dot{c}_0&=&-\frac{2\gamma_2\Omega'^{2}_1}{\gamma_1\gamma_2+4\Omega_{\mathrm{e}}^2}c_0+i\frac{4g\Omega_{\mathrm{e}}\Omega'_1}{\gamma_1\gamma_2+4\Omega_{\mathrm{e}}^2}c_2 \\
\dot{c}_2&=&-\left(\frac{2g^2\gamma_1}{\gamma_1\gamma_2+4\Omega_{\mathrm{e}}^2}+\frac{\kappa+\gamma_3}{2}\right)c_2+i\frac{4g\Omega_{\mathrm{e}}\Omega'_1}{\gamma_1\gamma_2+4\Omega_{\mathrm{e}}^2}c_0,
\end{eqnarray}
where we have assumed for simplicity that all couplings are real. Solving for the $c_2$ amplitude gives $c_2(t)=i\frac{\alpha}{\beta}e^{-\lambda t}\sinh(\beta t)$, where we have defined
\begin{eqnarray}
\alpha&=&\frac{4g\Omega_{\mathrm{e}}\Omega'_1}{\gamma_1\gamma_2+\Omega_{\mathrm{e}}^2} \\
\beta&=&\frac{1}{4}\frac{\sqrt{((\gamma_1\gamma_2+4\Omega_{\mathrm{e}}^2)(\kappa+\gamma_3)-4\gamma_2\Omega'^{2}_1+4g^2\gamma_1)^2-256g^2\Omega'^{2}_1\Omega_{\mathrm{e}}^2}}{\gamma_1\gamma_2+4\Omega_{\mathrm{e}}^2} \\
\lambda&=&\frac{g^2\gamma_1+\Omega'^{2}_1\gamma_2}{\gamma_1\gamma_2+4\Omega_{\mathrm{e}}^2}+\frac{\kappa+\gamma_3}{4}. 
\end{eqnarray}

The population, $\rho_{e3,1}(t_1)$, in state $\ket{\textrm{e}_3}\ket{1_{\text{ph}}}$ at time $t_1$ is $\abs{c_2(t_1)}^2$ while the population, $\rho_{e3,0}(t_1)$, in state $\ket{\textrm{e}_3}\ket{0_{\text{ph}}}$ at time $t_1$ is
\begin{equation} \label{eq:integration}
\rho_{e3,0}(t_1)=\kappa\int_0^{t_1}\abs{c_2(t)}^2e^{-\gamma_3(t_1-t)}dt. 
\end{equation}
At time $t_1$ the first laser ($\Omega_1$) is switched off and the second laser $\Omega_2$ is turned for a time $t_2-t_1$ to transfer any population in the state $\ket{\textrm{e}_3}$ to state $\ket{0}$. Note that we choose $t_1$ long enough to ensure that the population in $\ket{0}$ is negligible at time $t_1$. Assuming that $\Omega_2\gg\gamma_3$, we can neglect any spontaneous emission during the transfer 
and estimate the resulting population, $\rho_{0,\lambda}$ in state $\ket{0}\ket{\lambda_{\mathrm{E}}}$ at the end of the sequence as
\begin{equation}
\rho_{0,\lambda}\approx\frac{\kappa_{\mathrm{f}}}{\kappa}(\rho_{e3,0}(t_1)+\rho_{e3,1}(t_1)).
\end{equation}

The above sequence is repeated after the populations in the qubit states $\ket{0}$ and $\ket{1}$ has been flipped in order to generate the spin-photon entangled target state. Due to this symmetric generation scheme, we can estimate the overall fidelity of the generated spin-photon state with the target state as $F\approx\kappa\rho_{0,\lambda}/\kappa_{\mathrm{f}}$, with $\rho_{0,\lambda}$ as defined above. Here, we have assumed that if the photon is lost (with rate $\kappa_l$), this is a detectable error. Heralding on a successful outcoupling to the fiber thus gives a success probability of $\kappa_{\mathrm{f}}/\kappa$. In reality, spontaneous emission, from e.g. the $\ket{\textrm{e}_2}$ and $\ket{\textrm{e}_1}$ states, may also result in the absence of a telecom photon, but to obtain a lower bound on the fidelity, we have assumed that such errors are not detectable. 

From Eq.~(\ref{eq:integration}), we find that
\begin{align}
F&\approx\frac{\kappa\abs{\alpha}^2}{4\abs{\beta}^2}\Big(\frac{e^{-2(\lambda-i\Im{\beta})t_1}}{2(\lambda\!+\!i\Im{\beta})\!-\!\gamma_3}\!+\!\frac{e^{-2(\lambda+i\Im{\beta})t_1}}{2(\lambda\!-\!i\Im{\beta})\!-\!\gamma_3}\!-\!\frac{e^{-2(\lambda-\Re{\beta})t_1}}{2(\lambda\!-\!\Re{\beta})\!-\!\gamma_3} \nonumber \\
&\quad-\!\frac{e^{-2(\lambda+\Re{\beta})t_1}}{2(\lambda\!+\!\Re{\beta})\!-\!\gamma_3}\!+\!\frac{8\abs{\beta}^2(2\lambda-\gamma_3)e^{-\gamma_3t_1}}{(\beta^2\!-\!(2\lambda\!-\!\gamma_3)^2)^2\!-\!2(\beta^2+(2\lambda\!-\!\gamma_3)^2)(\beta^*)^2+(\beta^*)^4}\Big)\nonumber \\
&\quad+\frac{\abs{\alpha}^2}{\abs{\beta}^2}e^{-2\lambda t_1}\abs{\sinh(\beta t_1)}^2.
\end{align}
Here $\Im{}$ ($\Re{}$) is the imaginary (real) part and $*$ is the complex conjugate. In the limit $\kappa\gg g, \Omega_1, \Omega_{\mathrm{e}},\gamma$, the above expression simplifies to 
\begin{align}
F&\approx\frac{64g^2\kappa\Omega'^{2}_1\Omega_{\mathrm{e}}^2}{(4g^2\gamma_1\!+\!\gamma_1\gamma_2\kappa\!+\!4\kappa\Omega^2_e)((4g^2\!+\!\gamma_2\kappa)(\gamma_1\gamma_3\!-\!4\Omega'^2_1)\!+\!4\gamma_3\kappa\Omega^2_e)}\nonumber \\
&\quad\Big(e^{-\frac{4(4g^2+\gamma_2\kappa)\Omega'^{2}_1}{4g^2\gamma_1+\gamma_1\gamma_2\kappa+4\kappa\Omega^2_e}t_1}-e^{-\gamma_3t_1}\big).    
\end{align}

Defining the cooperativity $C=\frac{g^2}{\kappa(\gamma_2+\gamma_3)}$ and choosing $\Omega'_1=aC\gamma_1$, $\Omega_{\mathrm{e}}=C\gamma_2$, and $\gamma_1t_1=\ln C/C$, we find an error scaling as $1-F\propto\ln C/C$ in the limit $C\gg1$ as long as $4a^2\geq\gamma_2/(\gamma_2+\gamma_3)$. Note that we need $a\lesssim 1$ for our adiabatic elimination to be valid. While the analytical expression provides insight on how the parameters should scale to suppress the error, we also optimize the driving strengths ($\Omega$) and the time ($t_1$) in the numerical simulations to find the minimal error.

\section{Alternate Implementations} \label{ap:level scheme}
\begin{figure}
    \centering
    \includegraphics[width = 14cm]{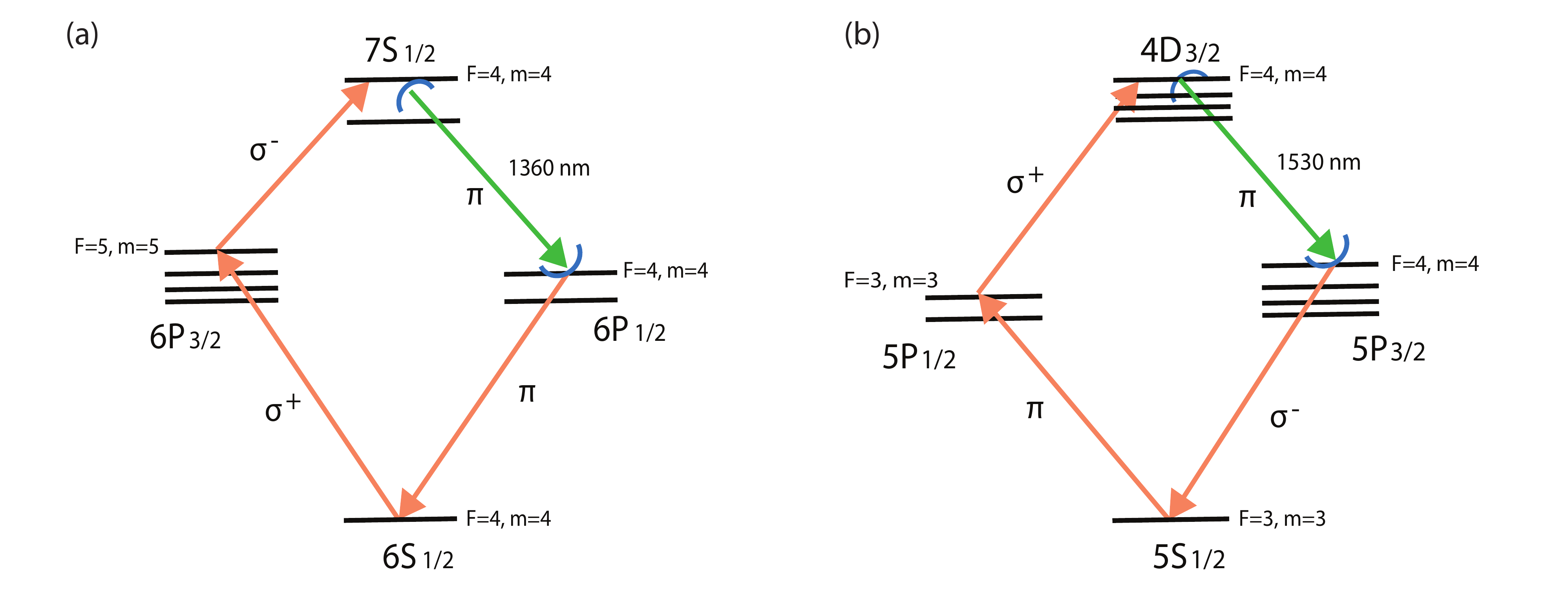}
    \caption{(a) A level scheme based on stretched states of cesium. Here the $\sigma^+$ transition is sensitive to polarization impurities, the $\sigma^-$ transition is insensitive to both $\pi$ and $\sigma^+$ polarizations, the cavity transition is insensitive to any $\sigma^+$ polarizations, and the final $\pi$ transition is insensitive to any $\sigma^+$ or $\pi$ polarizations. (b) A possible level scheme involving the rubidium atom with stretched states. This scheme would allow attainment of telecom photons at 1530 nm}
    \label{app:Stretch-rubidium}
\end{figure}

The entanglement scheme presented in the main text is very general and suitable atomic level schemes can be found for several atomic species and multiple choices of sub-levels. For cesium, apart from the clock state implementation presented in the main text, one can also find a level scheme based on stretched states, where off-resonant transitions are forbidden for pure polarizations. Schemes based on stretched levels have an additional advantage of being insensitive to a subset of polarization impurities based on the level selections (figure~\ref{app:Stretch-rubidium}(a)). However, the coherence time of stretched state qubits are smaller as they are more sensitive to magnetic field fluctuations. Moreover, we find that for the given waveguide design the polarization of $\sigma^+$, $\sigma^-$ fields are not well maintained.

The generalised implementation shown in figure~\ref{fig:figure1}(b) can be extended to other systems with a telecom transition from the first excited states. A possible implementation of the scheme in rubidium, another widely studied atomic system, is shown in figure~\ref{app:Stretch-rubidium}(b). Rubidium gives a selection of four different telecom transitions based on the choice of excitation to levels $4D_{3/2}$ or $6S_{1/2}$ from its two first excited states $5P_{1/2}$ and $5P_{3/2}$ respectively. The implementation shown here makes use of the telecom transition between states $4D_{3/2}$ and $5P_{3/2}$ at a wavelength of 1530~nm. This transition in the telecom C band allows low loss fiber propagation along with the ability to couple to telecom nodes of other memory platforms such as erbium doped solid state crystal memories. A scheme similar to figure~\ref{fig3}(a), making use of the magnetic insensitive $\mathrm{\mathrm{m_F}}=0$ state can be implemented in rubidium for both $4D_{3/2}$ and $6S_{1/2}$. However, the hyperfine splittings of rubidium are smaller compared to cesium making it more sensitive to polarization impurities.\\

\end{document}